\documentclass[12pt]{article}
\pdfoutput=1

\usepackage{scicite}
\usepackage{graphicx}

\newenvironment{sciabstract}{%
\begin{quote} \bf}
{\end{quote}}

\newcounter{lastnote}

\title{A photoinduced metallic phase of monoclinic vanadium dioxide.}

\author
{Vance R. Morrison$^{1}$, Robert. P. Chatelain$^{1}$, Kunal L.
Tiwari$^{1}$,\\
 Ali Hendaoui$^{2}$, Andrew Bruh\'{a}cs$^{1}$, Mohamed Chaker$^{2}$,
 Bradley
 J. Siwick$^{1\ast}$\\
\\
\normalsize{$^{1}$Departments of Physics and Chemistry, Center for the
Physics of Materials, McGill University}\\
\normalsize{801 Sherbrooke St. West, Montreal, QC Canada}\\
\normalsize{$^{2}$INRS EMT, Varennes, QC, J3X 1S2, Canada}\\
\\
\normalsize{$^\ast$To whom correspondence should be addressed; E-mail:
bradley.siwick@mcgill.ca.}
}

\date{}

\begin{document}

\baselineskip20pt

\maketitle

\begin{sciabstract}
The complex interplay between several active degrees of freedom (charge, lattice, orbtial and spin order) is thought to determine the electronic properties of
many oxides, but the respective role of the various contributions is often
extremely difficult to determine.  Vanadium dioxide (VO$_2$) is a
particularly notorious example. Here we report on combined ultrafast
electron
diffraction (UED) and infrared transmissivity experiments in which we
directly watch and separate the lattice and charge density reorganizations
that are associated with the optically-induced semiconductor-metal
transition
(SMT) in VO$_2$.  These studies have uncovered a previously unreported
photoinduced
transition to a metastable state with the periodic lattice
distortion (PLD) characteristic of the insulator intact, but differing by a
1D rearrangement of charge density along the octahedrally coordinated
vanadium dimer chains and a transition to metal-like mid IR optical
properties.  The results demonstrate that UED is capable of
following details of both lattice and electronic structural dynamics on the
ultrafast timescale.
\end{sciabstract}

Understanding the pathways through which microscopic interactions lead to the emergent properties of materials is the central problem of condensed matter physics. 
Collective phases affected by {\it multiple} collaborating or competing interactions
\cite{Imada1998,Dagotto2005,Lee2006}
have provided a great challenge in this respect.  The development of new
methods
(both experimental and theoretical) that are able to parse the mechanistic
role(s) of several
active interacting degrees of freedom are essential to developing an understanding of the properties
of such phases.
In this article we demonstrate that dramatic recent improvements in
ultrafast electron diffraction
instrumentation\cite{Siwick2003,Ruan2007,Sciaini2011,Gao2013}
provides such a capability by addressing the nature of the much debated semiconductor to metal
transition (SMT) in VO$_2$ \cite{Whittaker2011}.

At approximately 343 K, VO$_2$ undergoes a first order transition between
two
crystalline phases (Fig. 1A).  This structural phase transition (SPT) is
accompanied by a dramatic
change in conductivity; as much as 5 orders of magnitude in high quality
single crystals\cite{Berglund1969a}.
The high temperature phase, Figure 1A left, is metallic with a rutile
crystalline structure ({\it R}, P4$_2$/mnm).
The low temperature phase, Figure 1A right, is characterized by
semiconducting electronic behavior (E$_g$ $\sim$0.6 eV) and monoclinic
structure (M1,
P2$_1$/c).
The SPT may be understood roughly as the advent of a charge density wave
along
the rutile $c$-axis with wave vector $2\vec{c}_R$, which leads to a doubling
of
unit cell along this direction. This periodic lattice distortion (PLD)
dimerizes vanadium atoms along the $\vec{c}_R$ direction, spaced by 2.85
\AA{} in the high temperature phase, into alternating V-V separations of
2.62
\AA{} and 3.16 \AA{}. The dimers are also rotated slightly with respect to
$\vec{c}_R$.

A long-standing challenge to understanding this
SMT has been
to determine the relative role of electron-lattice interactions (lattice and charge order) and electron-electron interactions (dynamical correlations and orbital selection)
to the change in properties and the nature of the semiconducting
phase\cite{Goodenough1971,Zylbersztejn1975,Wentzcovitch1994,Eyert2002,Biermann2005,Weber2012}.
Here we address this question directly by making use of the
orders-of-magnitude beam brightness enhancement provided by
radio-frequency compressed ultrafast electron diffraction\cite{Chatelain2012}
in combination with time-resolved
IR transmittance measurements to interrogate both structure and electronic
properties. The combined
approach makes it possible to map the reorganization of the VO$_2$ unit cell
{\it during} the optically induced
transition\cite{Cavalleri2004,Wall2013} in unprecedented detail while
simultaneously determining electronic properties.  The results unambiguously
demonstrate a photoinduced transformation to a long-lived state with
metal-like mid IR optical properties and the PLD
(or charge density wave order) of the semiconducting M1 phase intact.  This state
differs from the
equilibrium rutile metal crystallographically, and in that it only involves a 1D reorganization of
charge density rather than a transition
to the isotropic 3D electronic state of the high temperature
phase\cite{Koethe2006,Haverkort2005}.

In these experiments pulsed-laser deposition grown
polycrystalline VO$_2$ films (Figs. 1B, 1C)\cite{Hendaoui2013}, initially in
the
low temperature 
M1 phase (at $\sim$310K), are subject to optical (800nm) excitation
with 35
fs laser pulses.  The time-dependence of the changes in structure and
electronic properties following optical excitation of the material are
determined using pump-probe UED and time resolved IR transmittance
measurements\cite{Supporting2014}.  The time-resolved transmission electron
powder diffraction
data obtained using UED (Fig. 2) provides an extremely rich and
detailed
view
of the structural dynamics following photoexcitation. Raw (background
subtracted) UED data for 20 ps following photoexcitation at an intermediate
pump fluence of 20 mJ/cm$^2$ is shown in Fig. 2A. Of particular note are
several weak
reflections (30\={2}, 12\={2} and 31\={3}, indicated by red lines), which
are
allowed in the M1 phase due to the PLD and the doubling of the unit
cell along $\vec{c}_R$, but not in the {\it R}
phase.  The intensity of these peaks clearly decreases with time following
photoexcitation, which is indicative of the optically induced SPT that occurs in some of the sample at this
fluence.   Similar observations were made in previous ultrafast structural
measurements on VO$_2$ which identified a sub 500 fs timescale for aspects of
the
SPT\cite{Cavalleri2001,Baum2007}.  Beyond these specific peaks, however,
photoexcitation induces changes in diffracted intensity over the entire
scattering vector range shown.  This is clearly evident in Fig. 2B where the
time-dependent difference in diffracted intensity with respect to
negative pump-probe delays (i.e. before photoexcitation) is plotted. The
presence of multiple time scales can be seen clearly in Fig. 2C which shows
the
time dependent intensity of several diffraction features labeled in Fig. 2A.
After photoexcitation at 20 mJ/cm$^2$ there is a fast,
310$\pm$160 fs, decrease in the intensity of diffraction peaks associated
with the PLD
(e.g., 30\={2}), followed by a slow
1.6$\pm$0.2 ps time constant increase in the intensity of most peaks in the
range $s < 0.45$ \AA$^{-1}$ that are present for both phases (e.g., the 220
and 200
features).  These are the only two ultrafast time
constants observed in the data up to 10 ps.

The amplitude of these two
qualitatively distinct diffraction signatures each scale linearly with
fluence, but have different slopes and threshold fluences: 9 mJ/cm$^2$
 for fast pump-induced changes to peaks associated with the PLD in the 
 M1 phase and 2 mJ/cm$^2$ for
 the slow changes.  At pump fluences
below $\sim$9 mJ/cm$^2$ (hatched region in Fig. 2D) there is no change to the
intensity of diffraction peaks associated with the PLD in the M1 phase.
Inset in Fig. 2D are
time-resolved IR transmittance curves at 5 microns (0.25 eV) for the VO$_2$
film at several
pump fluences
below 9 mJ/cm$^2$.  These curves
show a persistent decrease in IR transmissivity that increases
with pump fluence and reaches an amplitude of  $>$99\% by 3.7 mJ/cm$^2$.
This
observation is in quantitative agreement with previous experiments using
multi-THz spectroscopy on pulsed laser deposition grown VO$_2$ films
\cite{Pashkin2011a,Cocker2012}, and is indicative of
a (partial) transition to a state with metallic-like AC conductivity at a
threshold pump-fluence of approximately 2 mJ/cm$^2$.

We isolate the full spectrum of
diffracted intensity changes that correspond to the fast and slow components
 in the time-domain by choosing reference time points
for computing the intensity differences that separate these dynamics, i.e.,
$t = -1$ ps in Fig. 2E (fast dynamics) and $t  = 2$ ps
in
Fig. 2F (slow dynamics).  It is important to note that unlike the fast
dynamics in Fig. 2E,
the slow dynamics are dominated by increases in peak intensity over a limited
range
of scattering vector (s $<$ 0.45 \AA$^{-1}$)
for which electron scattering is known to be particularly sensitive to the
valence charge
distribution\cite{Zuo2004,Zheng2005}.
In addition, these slow dynamics are absent for reflections whose reciprocal
lattice vector is perpendicular to $\vec{c}_R$ ($h_M=0$), as indicated with
grey dotted lines in Fig. 2.  This observation unambiguously
establishes that electron structure factors
orthogonal to $\vec{c}_{\textrm{R}}$ are largely unaffected by the slow
process; i.e.
the slow process corresponds to a 1D modification of the electrostatic
crystal potential in
the octahedrally coordinated vanadium chains oriented along
$\vec{c}_{\textrm{R}}$.

The two diffraction signatures described above represent qualitatively
distinct structural reorganizations within VO$_2$ following
photoexcitation. This can be understood by computing the pump-induced changes
to the radial
pair distribution function (PDF) \cite{Abeykoon2012} directly from the
observed
changes in diffracted intensity shown in Fig. 2E and F.  These curves,
shown in Fig. 3A and B, represent the time-dependent difference in the radial
autocorrelation function of the crystal
potential with respect to the reference time point.
The computed difference PDF for the
fast dynamics is shown in Fig. 3A and provides a straight forward structural
interpretation for this signal. The positive growing feature (II) corresponds
to
increased correlation at the {\it R} phase V-V bond length, 2.85 \AA,
while the adjacent negative going features (I, III) represent a
reduction at the dimer (2.62 \AA) and unpaired (3.16
\AA) distances of the {\it M1} phase as indicated in Fig. 1.  Thus, the fast
dynamics correspond
 to non-thermal melting of the PLD in a fluence dependent fraction of crystallites.  In these crystallites the vanadium atomic positions relax to their equilibrium
 {\it R}-phase separation on the 300 fs timescale.  Extrapolating the linear
 scaling of these dynamics with fluence indicates that $\sim$43
 mJ/cm$^2$ is required to melt the PLD (CDW order) in the entire film. Pump fluences
{\textless}9 mJ/cm$^2$ are insufficient to initiate this non-thermal SPT in any crystallites/volume, and
leave the PLD and M1 crystal structure completely intact. Below this $\sim$9
mJ/cm$^2$ threshold only the slow dynamics is observed (Figs. 2F and S4).
The diffraction signature of these dynamics is identical below and above the SPT threshold, demonstrating that the slow and fast components represent distinct transitions occuring in different crystallites/volumes due to the heterogeneity of these pulsed laser deposition samples.

In contrast to the above, the slower dynamics do not correspond to a
structural rearrangement of the lattice (which
 result in a conservation of diffracted intensity like that seen for the fast
 dynamics in Fig 2E).
 The difference PDF (Fig. 3B) for the slow changes is dominated by negative
 going
features at 1.3 \AA{} (IV) and 4.4 \AA{} (VI), equal to half the V-V dimer
bondlength and the undimerized V-V
separation plus half the V-V dimer bondlength respectively (Fig. 1A).
Positive going changes are also observed at around 1.9 \AA{} (V), the
average V-O separation in the octahedron, and at \textless 0.8
\AA.  These observations are consistent with a collective
reorganization of valence charge density in the M1 phase that increases the
electron density in the vanadium dimers bonds
while decreasing the electrostatic potential on primarily the oxygen
atoms;  an effective modification of the atomic scattering factors.

Previous theoretical work on vanadium dioxide has focused
on the behaviour and occupancy of the three
bands formed from hybridized V-3\emph{d}/O-2\emph{p} states of
$t_{{2g}}
$ symmetry (Fig 4A) as the determining factor in the electronic properties of
VO$_2$ \cite{Goodenough1971,Eyert2002,Biermann2005,Weber2012}.
The orientation of the localized \emph{d}-orbitals from which these bands are
formed are
shown in Fig. 4B.  The $d_{xy}$ (also referred to as d$_\parallel$) and
$d_{xz}$ orbitals mediate sigma and pi type interactions between vanadiums
along
$\vec{c}_{\textrm{R}}$ respectively.  The $d_{yz}$ orbital is oriented
orthogonal
to $\vec{c}_{\textrm{R}}$.
There is broad agreement in the high temperature phase these three bands
almost completely overlap at
the
Fermi level\cite{Eyert2002,Biermann2005}, as shown in Fig. 4A-i.  This
results in roughly equal occupancy in
these bands and a nearly isotropic electronic state
\cite{Koethe2006,Haverkort2005}.  It has been suggested that the PLD in the
{\it M1} phase
splits the $d_{\parallel}$ states into bonding and
anti-bonding combinations sufficiently to open an insulating gap
\cite{Goodenough1971},
but density functional theory calculations using the local density
approximation maintain significant
density of states at the Fermi level in the M1 phase
\cite{Eyert2002,Biermann2005} as shown in Fig.
4A-ii\cite{Belozerov2012,Sakuma2009}

Recent work using cluster dynamical mean field theory points to
dynamical electron-electron correlations
acting in collaboration with the PLD as being responsible for the insulating
properties of the M1 phase (Fig.
4)\cite{Weber2012,Biermann2005}.  The results presented here support this
view.  We have demonstrated that optical excitation can induce a
long lived state with IR transmissivity like that of the metallic phase
(i.e., collapse of the optical bandgap to below 0.25 ev) in the absence of a
SPT.  This state has the PLD of
the insulating phase intact, but the valence charge distribution
significantly altered.
  The nature of the changes in charge density can be understood
from the symmetry of the changes in diffraction presented in Fig. 2F,
the orbitals shown in Fig. 4B and the difference PDF in Fig. 3B.  The
negative going feature at 1.3
\AA{} and 4.4 \AA{}
in Fig. 3B suggests an increase in the filling of the $d_{xy}$ subshell that
the mediate V-V dimer bonds.
The positive going features at 0.8 \AA{} and 1.9 \AA{} suggest reduced
filling
of the  $d_{xz}$ subshell, which reduces charge density on the
V and O atoms in the octahedral chains. The  $d_{yz}$
states oriented orthogonal to $\vec{c}_{\textrm{R}}$, which are understood to
be unoccupied in semiconducting VO$_2$  \cite{Haverkort2005, Weber2012},
remain unchanged.
Thus, optical excitation with fluences below the threshold required to melt the PLD drives a 1D
redistribution of occupancy in the d$_{xy}$ and d$_{xz}$ subshells,
not a transformation to the isotropic state of the equilibrium metal.
Suppression of correlation
induced splitting into upper (UHB) and lower (LHB) Hubbard bands, either
preferentially
in the $d_{xz}$ band (Fig. 4A iv) or in both $d_{xz}$ and $d_{xy}$ shells
(Fig. 4A ii), could lead to such a reorganization.  The first case
represents an optically induced orbital selective transition with a mixture
of localized (d$_{xy}$) and itinerant (d$_{xz}$) behaviour\cite{Yao2013}.
The picosecond timescale of
this transformation in addition to its long-lived nature suggests that the increased vibrational
excitation of the lattice due to carrier relaxation (e-ph coupling) is a
key factor in both inducing and maintaining the reorganization. The
nonequilibrium population of excited carriers relax
in $\sim$1ps in thin pulsed laser deposition grown VO$_2$ films
\cite{Pashkin2011a}.  However, the present measurements cannot rule out the possibility of other mechanisms affecting this stability, including kinetic trappingof the valence charge reorganization.
Here a tentative connection can be made to earlier work which identified metal-like phases of VO$_2$ with properties unlike that of the rutile, high-temperature metal.  Qazilbash \emph{et al.} observed
the emergence of nanoscale correlated metallic
domains with electronic properties unlike those of the rutile metal near the
transition
temperature\cite{Qazilbash2011,Qazilbash2007}.  The lattice structure of
these puddles was not identified.  Additionally, the work of Nag \emph{et al.} demonstrated that the SMT and SPT can occur non-congruently when initiated thermally, suggesting the presence of a metal-like, M1 phase of VO$_2$\cite{Nag2012}. The correlated metallic state observed in the thermally activated phase transition and the M1 phase metastable state accessed optically here may be related.

The results presented here have several important implications to our understanding of the hierarchy of roles for electron-lattice and electron-electron degrees of freedom in determining the properties of semiconducting VO$_2$.  
The profound decoupling of the SMT in mid IR optical properties and the SPT induced through optical excitation 
indicates that the PLD of the M1 phase is insufficient to fully explain the semiconducting gap.
 From the perspective of the
dramatic change in electronic properties, the principle role of the PLD
is to alter the accessibility of the bands formed by states of $d_{xy}$
symmetry\cite{Goodenough1971}.  With the
PLD in place these states are depopulated, and the highest energy occupied
bands have
a 1D character and are susceptible to further electronic ordering.  The isotropic
electronic character
of the equilibrium rutile metal\cite{Haverkort2005} cannot be realized with
the PLD intact.
Finally, the large threshold excitation fluence for the SPT compared to that for the observed electronic reorganization demonstrates that the latent heat of the first order phase transition at $\sim$ 340K is dominated by the SPT rather than the electronic transition.

In summary, we have demonstrated that substantial advances in UED have provided the capability to simultaneously probe the dynamics of both lattice structure and charge density on the femtosecond time scale.  
This provided new insights on the SMT in VO$_2$, uncovering a novel optically accessible state. 
Based on this study, UED is now positioned to provide deep insights into the nature of other strongly correlated materials through the disparate concurrent responses of active degrees of freedom in the time domain. 
Further, our results have relevance to the study of the interplay between valence charge and lattice structure in molecular and materials chemistry.
\nocite{Funding2014}

\bibliographystyle{Science}

\begin{figure}[h!]
\centerline{\includegraphics[width=12cm]{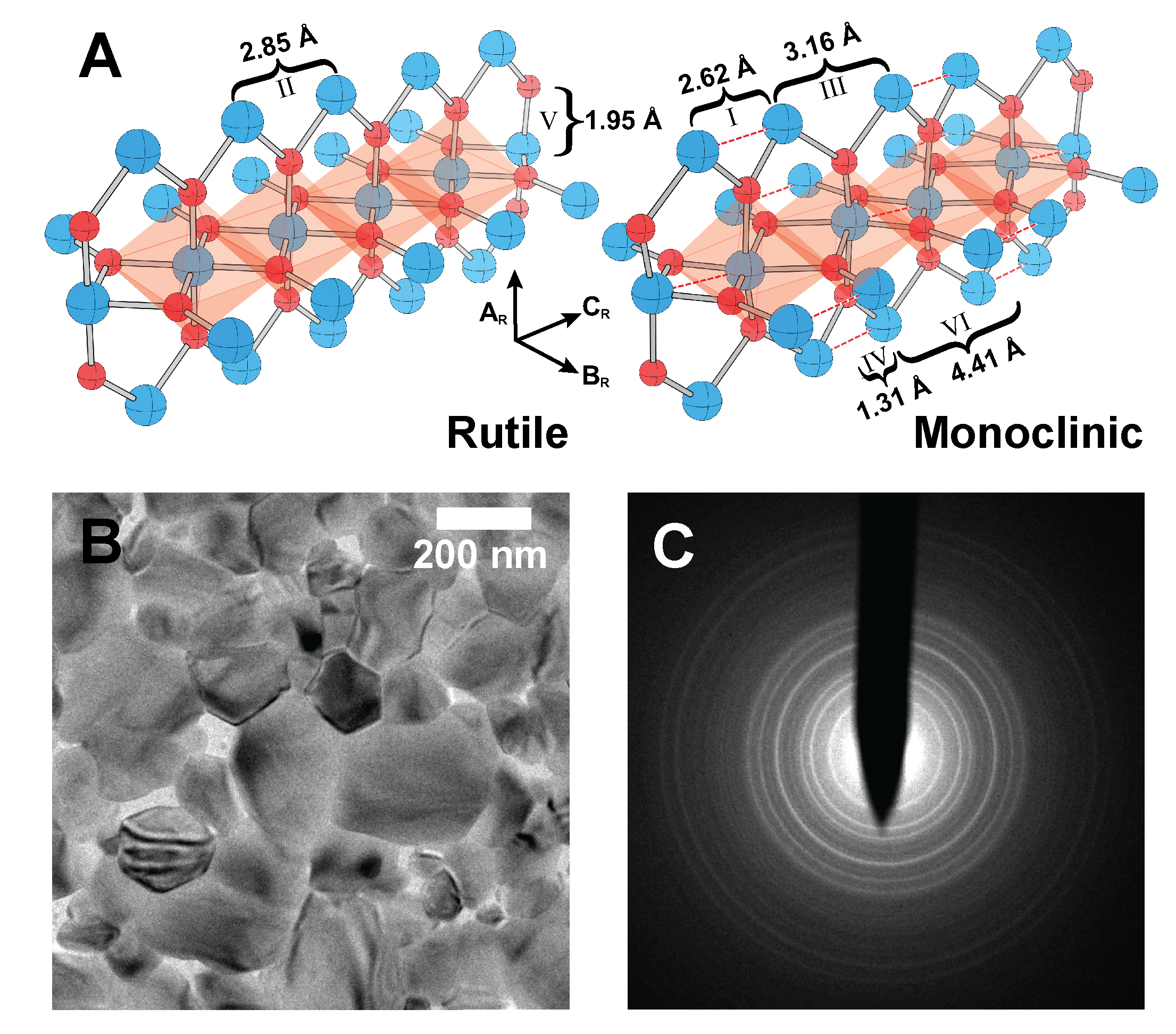}}
\caption{A) The structure of rutile VO$_2$ (left) and monoclinic
 VO$_2$ (right).  B) Transmission electron microscopy image of 
the pulsed laser deposition grown VO$_2$
sample used in these studies. C) Example electron powder diffraction pattern
of the monoclinic phase.}
\label{fig:fig1}
\end{figure}

\begin{figure}[htb]
\centerline{\includegraphics[width=16cm]{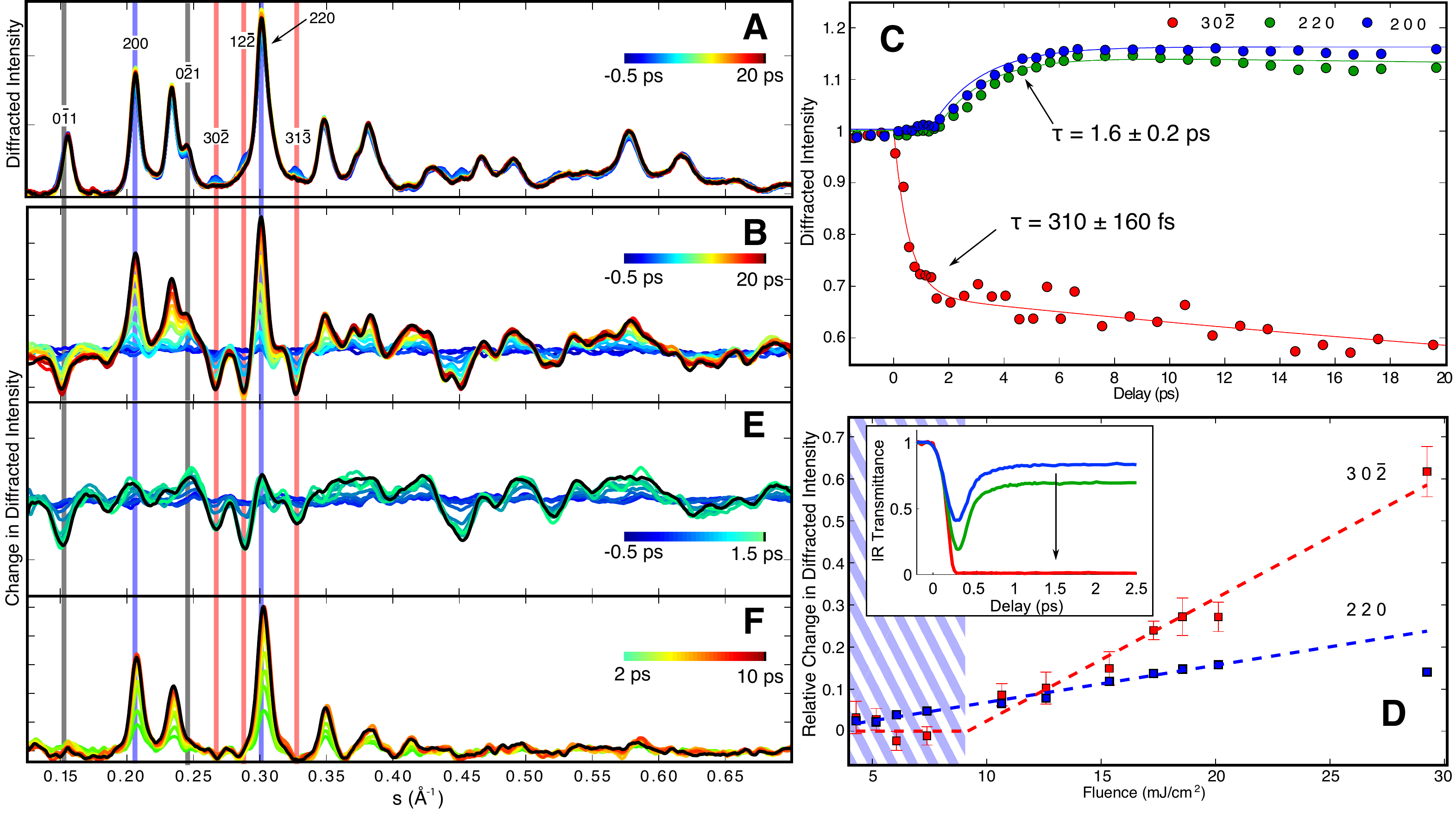}}
\caption{Structural dynamics during the SMT in VO$_2$.  A) Raw, background subtracted, UED data from 0-20 ps. Red vertical lines 
indicate several weak reflections allowed in the M1 phase due to the PLD but not in the R-phase.  
Blue lines indicate several peaks present in both equilibrium phases.  The grey lines indicate 
peaks for which $h_M=0$. B) Overall diffraction difference spectrum from
-0.5 to 20 ps, C) Time resolved diffraction peak intensity showing fast ($\sim$300fs) and slow ($\sim$1.6 ps) 
dynamics respectively for peaks indicated by red and blue vertical lines in the diffraction spectra
(leftmost panels). The dashed line demarks a delay of 1.5 ps and acts as guide to the eye. D)
Fluence dependence of the fast and slow signal amplitudes as measured for the (30\={2}) and (220) peaks
shown in C).  The range of fluences for which no SPT is observed is indicated by the hatched region.  
Inset) Time resolved IR (5 $\mu$m, 0.25 eV) transmissivity in this hatched fluence region (3.7 mJ/cm$^2$ (red), 2.7 mJ/cm$^2$ (green) 
and 1.9 mJ/cm$^2$ (blue)) display a persistent decrease to a very long-lived plateau ($>$100ps).  The amplitude of this
decrease reaches $>$99\% at 3.7 mJ/cm$^2$, indicating a significant closing of the semiconducting gap and a transition to a metallic-like state.     
E) Diffraction difference spectrum for the fast dynamics. The change in diffracted intensity from -0.5 ps to
1.5 ps and F) Diffraction difference spectrum for the slow dynamics.  The change in diffracted intensity from 2 ps to 10 ps (referenced to 2 ps) is shown. }
\label{fig:fig2}
\end{figure}

\begin{figure}[htb]
\centerline{\includegraphics[width= 8cm]{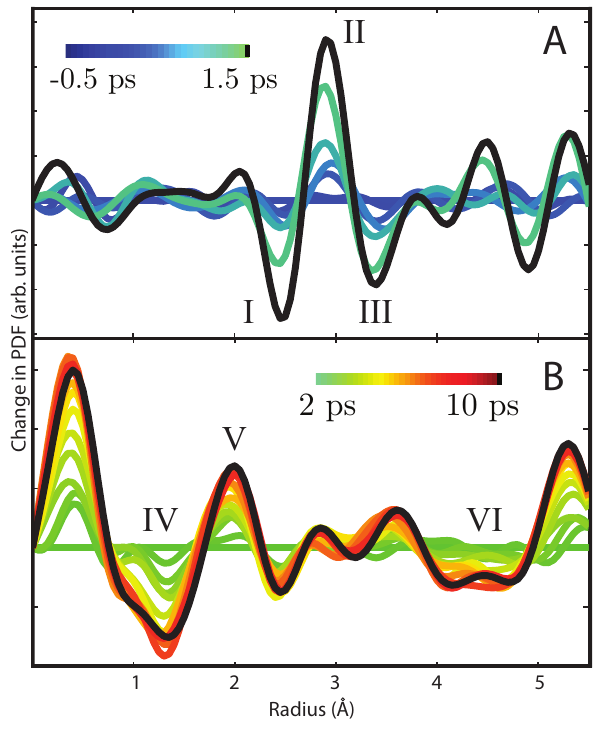}}
\caption{Difference pair distribution functions for the two observed characteristic time scales. A) Difference PDF from -0.5 ps to 1.5 ps referenced to -0.5 ps.  B)
Difference PDF from 2
ps to 10 ps, referenced to 2 ps.  The roman numerals correspond to the distances labeled in Fig 1.}
\label{fig:fig3}
\end{figure}

\begin{figure}[htb]
\centerline{\includegraphics[width=15cm]{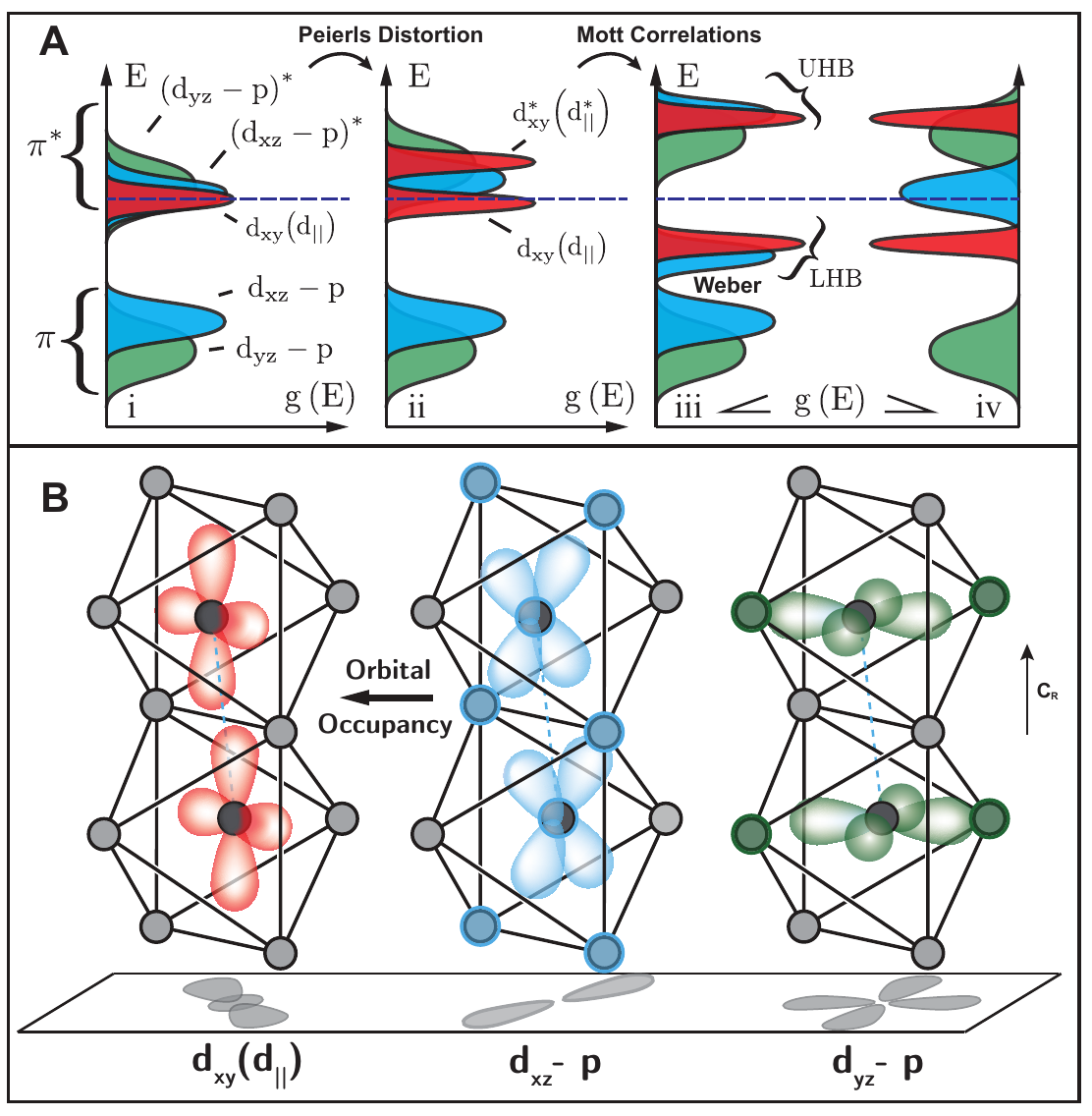}}
\caption{Effective band diagrams of VO$_2$ and illustrations of the associated molecular orbitals. A-i) Band diagram for the rutile, metallic phase. A-ii) Modified band diagram as a result of the PLD. A-iii) The effect of el-el correlations as described by Weber et al\cite{Weber2012}. A-iv) The band diagram resulting from the partial Mott melting of the $d_{xz}$ band. B) Illustrations of the $d_{xy}$, $d_{xz}$, and $d_{yz}$ molecular orbitals.}
\label{fig:fig4}
\end{figure}

\clearpage
\renewcommand{\thefigure}{S\arabic{figure}}
\section*{Supplementary Material}

\setcounter{figure}{0}  
The measurements presented here were made on a home built ultrafast electron diffractometer (Fig. \ref{fig:schematic}) employing radio-frequency pulse compression techniques\cite{VanOudheusden2007}. This approach has very recently been shown to offer dramatic improvements in instrument performance\cite{Chatelain2012,Gao2012}.
These techniques allow us to produce compressed electron pulses with $\sim$10$^6$ 
electrons per pulse with $\sim$300 fs time resolution\cite{Chatelain2012}.

In these experiments, VO$_2$ specimens initially in the {\it M1} phase ($\sim$310 K) are photoexcited with 35 fs, 1.55 eV (800 nm) laser 
pulses at near normal incidence ($\sim$10 degrees) with a range of excitation fluences; the evolution 
of the SMT was then probed after a variable delay time with an ultrashort electron pulse with 
$\sim$500,000 electrons per pulse.  These electron pulses were generated through photoemission in a DC, high voltage 
electron gun with 266 nm laser pulses and accelerating voltages up to 95 kV.   Although the diffractometer is capable of operating with a 
repetition rate of 1 kHz, a delay of 20 ms between pump laser pulses was required before reinitiating the transition in 
order to allow the samples to completely relax back to the insulating phase.   The diffraction pattern (Fig. \ref{fig:DiffPatt}) produced at each 
delay is the result of approximately 50 individual fifteen second exposures and was detected on a Gatan Ultrascan 1000, a phosphor coated CCD with near single electron detection capabilities.  

In addition to the ultrafast electron diffraction measurements performed, ultrafast infrared 
transmittance measurements were conducted under identical pump-probe conditions.  
The IR pulses (5 $\mu$m wavelength) were generated using difference frequency generation from the signal and idler beams of an IR optical parametric amplifier\cite{Kaindl2000}.  The ultrafast mid-IR transmission measurements were made using the femtosecond pulse acquisition spectrometer from Infrared Systems Development, based on 64 pixel Mercury Cadmuim Telluride IR detector arrays.  The VO$_2$ samples
were held at atmospheric pressure as opposed to the 10$^{-7}$ Torr used for the electron diffraction
experiments.

The high-quality stoichiometric polycrystalline vanadium dioxide films used in these measurements 
were synthesized at the Laboratory of Micro- and Nanofabrication facility at INRS by means of 
reactive pulsed laser deposition process. A pure (99.95\%) vanadium target was used with a KrF excimer laser ($\lambda$ = 248 nm) in an oxygen 
environment with a pressure of 15 mTorr and substrate temperature of 500 $^\circ$C\cite{Hendaoui2013,Hendaoui2013a}.  The samples were 70 nm thick and deposited on 50 nm thick amorphous silicon nitride windows.  Temperature dependent resistivity curves of the samples used are shown in Fig. \ref{fig:temp} 
and display the characteristic hysteretic behaviour of VO$_2$ with a hysteresis width of 10 K.

\clearpage

\begin{figure}[h!]
\centerline{\includegraphics[width=16cm]{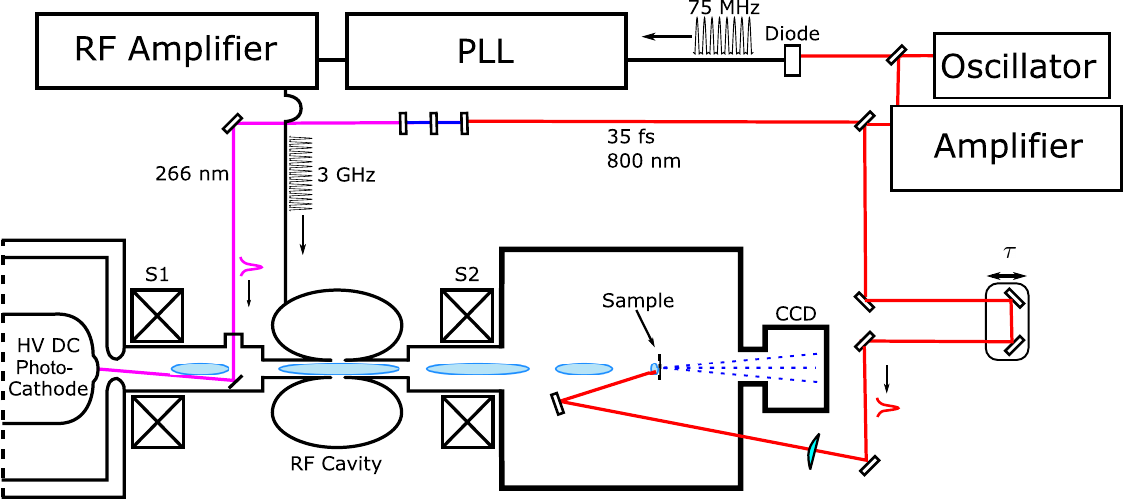}}
\caption{Schematic diagram of the ultrafast electron diffractometer including the synchronization electronics and the chirped pulse amplification laser system.  The sample is pumped with 35 fs, 800 nm laser pulses, while third harmonic generation is used to produce 266 nm, UV pulses.  A fast photo-diode is used to detect the timing of the pulses from the ultrafast oscillator; a phase-locked loop (PLL) is then used to produce synchronized RF pulses which are then amplified by an RF amplifier.  These pulses then drive the RF compression cavity. Two magnetic lenses, S1 and S2, are used to collimate the electron beam as it exits the electron gun, and then to focus the beam at the CCD detector.}
\label{fig:schematic}
\end{figure}

\begin{figure}[htb]
\centerline{\includegraphics[width=14cm]{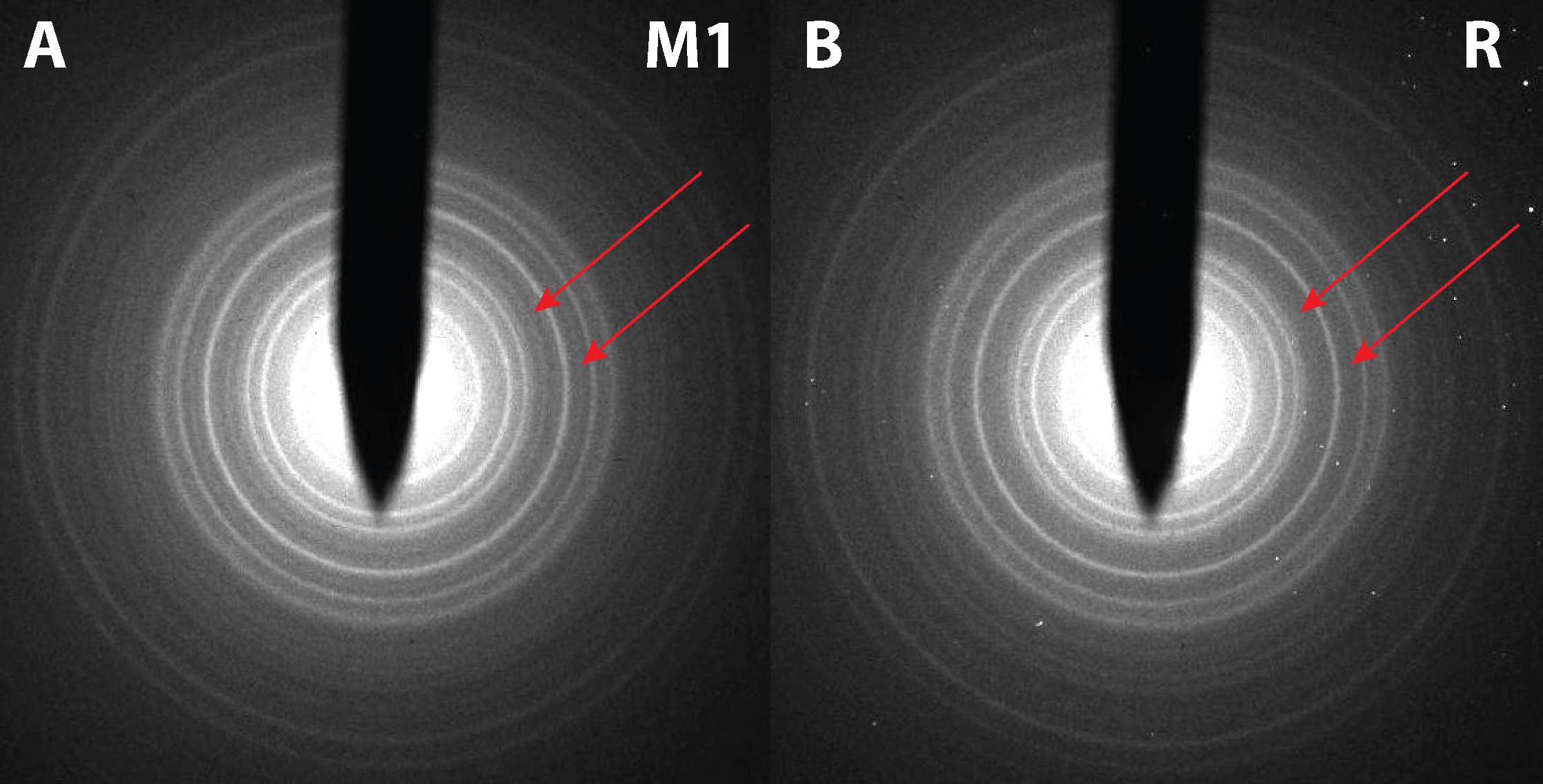}}
\caption{Diffraction patterns of low temperature (A) and high temperature (B) phases of VO$_2$.  The red arrows indicate the disappearance of diffraction features which are present in the M1 phase, but not allowed by symmetry in the R phase.  The shadow in the center of the images is a beam block used to prevent the main electron beam from saturating the CCD.}
\label{fig:DiffPatt}
\end{figure}

\begin{figure}[htb]
\centerline{\includegraphics[width=12cm]{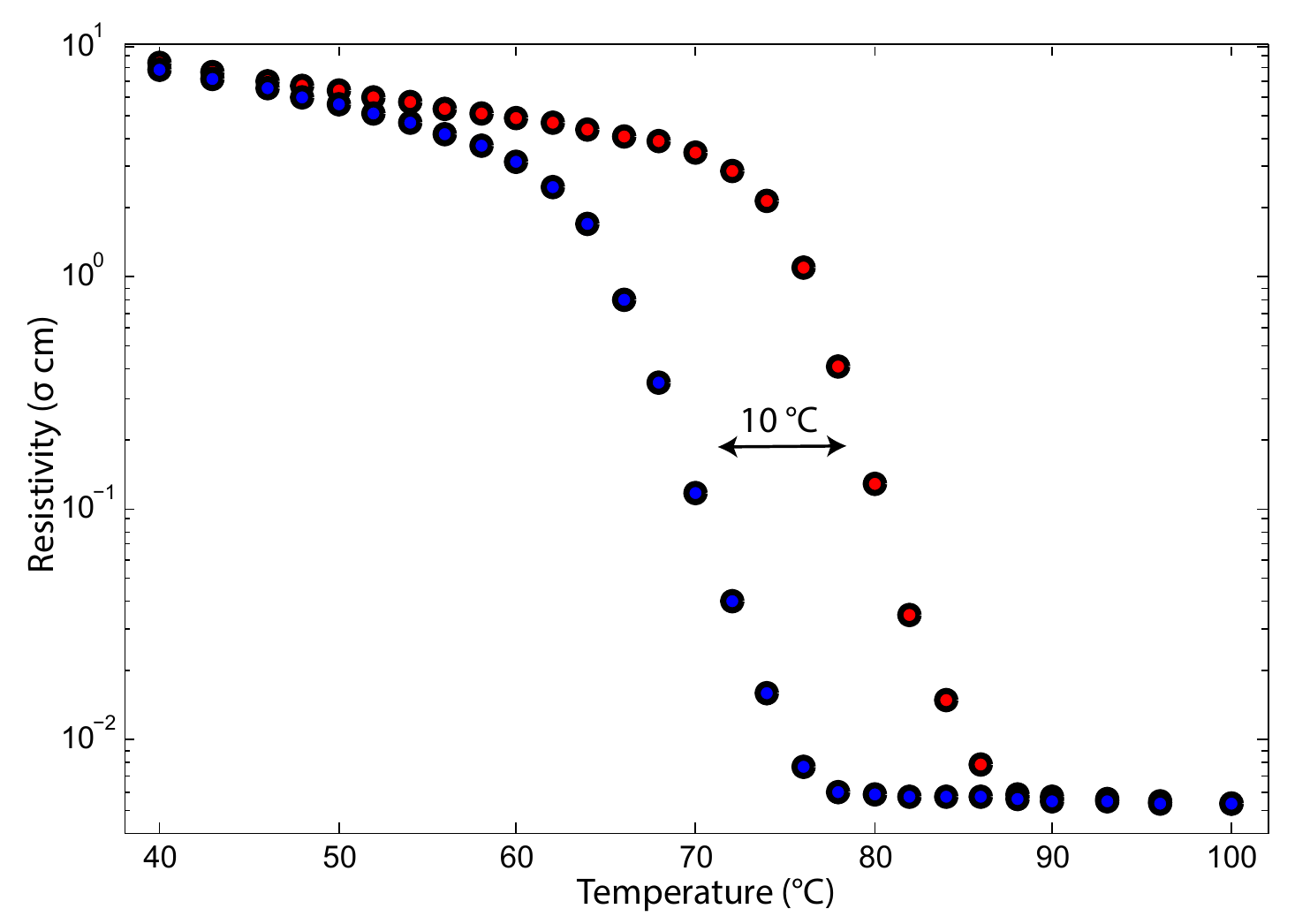}}
\caption{Temperature dependent resistivity curve of the IMT to in the thin film VO$_2$ samples 
used.  The hysteresis width is 10 $^{\circ}$C with increasing temperature shown in red and 
decreasing temperature shown in blue.}
\label{fig:temp}
\end{figure}

\begin{figure}[htb]
\centerline{\includegraphics[width=16cm]{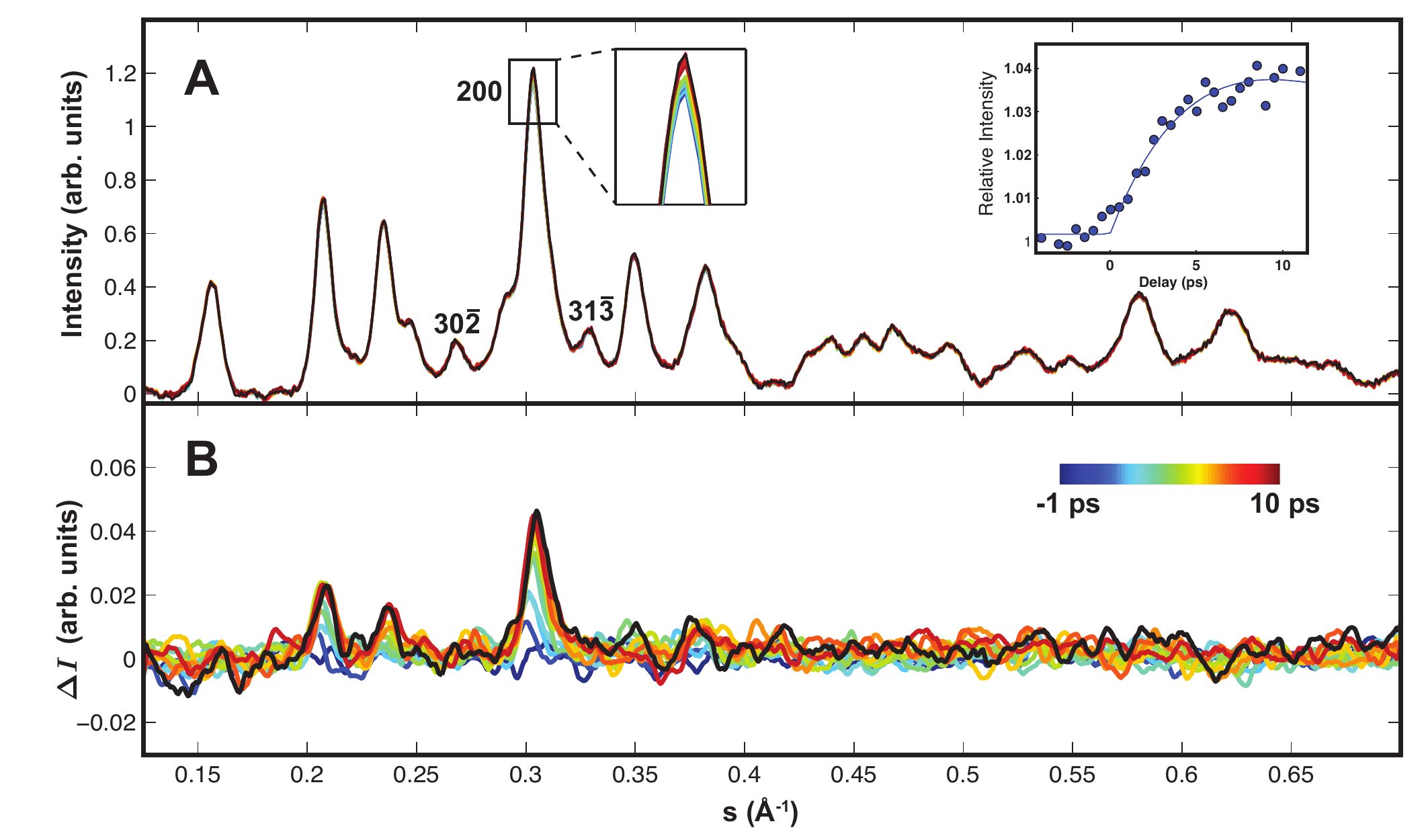}}
\caption{A) Time resolved, background subtracted, diffraction data from 0 ps to 10 ps and an excitation fluence of 6.1 mJ/cm$^2$. Inset) Time resolved intensity of the 220 reflection. B) Time resolved change in diffracted intensity shown in A). No decrease is seen in the 30\=2 and 31\=3 reflections, indicating that the PLD remains intact.  In addition, no changes are observed in the $s>0.5$ {\AA}$^{-1}$ region.}
\label{fig:LowFluence}
\end{figure}

\end{document}